\documentclass[review]{elsarticle}

\usepackage{lineno,hyperref}
\modulolinenumbers[5]

\journal{Journal of \LaTeX\ Templates}

\usepackage{graphicx}

\journal{C. R. Physique}

\begin{document}

\begin{frontmatter}



\title{Towards a Re-definition of the Second Based on Optical Atomic Clocks}


\author{Fritz Riehle}
\address{Physikalisch-Technische Bundesanstalt (PTB), Bundesallee 100, 38116 Braunschweig, Germany, Tel (+49) 531 592 4010, email: fritz.riehle@ptb.de}

\begin{abstract}
The rapid increase in accuracy and stability of optical atomic clocks compared to the caesium atomic clock as primary standard of time and frequency asks for a future re-definition of the second in the International System of Units (SI). The status of the optical clocks based on either single ions in radio-frequency traps or on neutral atoms stored in an optical lattice  is described with special emphasis of the current work at the Physikalisch-Technische Bundesanstalt (PTB). Besides  the development and operation of different optical clocks with estimated fractional uncertainties in the $10^{-18}$ range, the supporting work on ultra-stable lasers as core elements and the means to compare remote optical clocks via transportable standards, optical fibers, or transportable clocks is reported. Finally, the conditions, methods and next steps are discussed that are the prerequisites for a future re-definition of the second. 
\end{abstract}



\end{frontmatter}


\section{Introduction}
\label{Introduction}
\begin{figure}[htbp]
\centering\includegraphics[width=10cm]{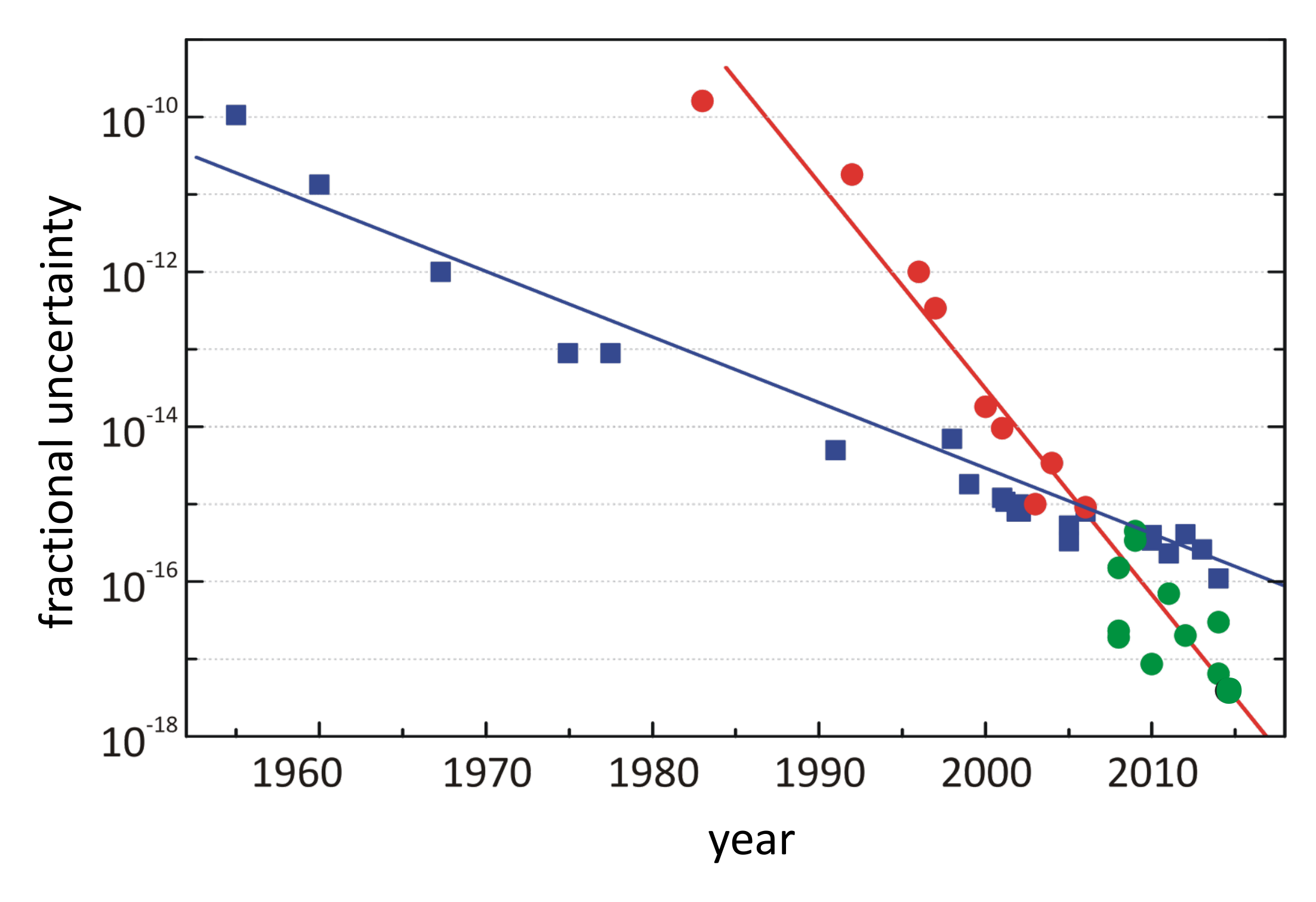}
\caption{\label{OptvsMicro} Fractional uncertainty of primary caesium atomic clocks to realize the second in the SI (squares) and optical frequency standards (dots). Green dots represent optical clocks with estimated standard uncertainties whose optical frequency can be determined only with the accuracy of the frequency comparison with a primary caesium clock.}
\end{figure}

Since 1967 the second as the base unit of time in the International System of Units (SI) is defined via an unperturbed  hyperfine transition in atomic caesium \cite{bip67} that is interrogated by a radio frequency of 9~192~631~770 Hz. In the past five decades the uncertainty to which the center of this hyperfine transition can be found in the best caesium atomic clocks could be reduced typically in each decade by an order of magnitude (Fig. \ref{OptvsMicro}). A few years ago, optical atomic clocks (see e.g. \cite{pol13,lud14}) where a laser is stabilized to a suitable optical transition have fulfilled their long standing promise to supersede caesium atomic clocks with respect to their lower frequency instability and their lower inaccuracy. Accuracy is here understood as the ability to reference the frequency of a clock to an unperturbed atomic or ionic transition. However, none of these ``better'' optical clock transitions with superior precision can realize the second more accurately than the best caesium atomic clock they have been compared to. It can therefore be anticipated that at some time the SI second will be re-defined via an optical clock transition \cite{gil11}. Already in the year 2001 the Consultative Committee for Time and Frequency (CCTF) discussed the situation and formed a working group that later became the Working Group on Frequency Standards (WGFS) under the auspices of CCTF and CCL (Consultative Committee for Length) to recommend {\it Secondary representations of the second} (whether optical or microwave), that could be used to realize the second in parallel to the caesium clock.  The establishment of these secondary representations of the second would furthermore help with the detailed evaluation of reproducibility at the highest level, and significantly aid the process of comparing different standards in the preparation of a future re-definition \cite{gil06b}. In 2006 the first five secondary representations (one in the microwave domain and four optical standards) were recommended by the International Committee of Weights and Measures (CIPM) on its $95^{th}$ session \cite{cip}. In 2012 the CIPM on its $101^{st}$ session updated and augmented that list \cite{cip} that can be found on the web site of the International Bureau of Weights and Measures (BIPM) \cite{bip}. 

The quantum projection noise limited instability of a clock that references an atomic transition is given by
\begin{equation}
\sigma_y(\tau) = \kappa \frac{\Delta \nu}{\nu} \sqrt{\frac{T_{int}}{N \tau}}
\label{instability}
\end{equation}
where $\kappa$ is a number of order unity that depends on the the type of interrogation, $\nu$ is the frequency of the optical clock, $\Delta \nu$ is the line width at the interrogation time $T_{int}$, $N$ is the number of absorbers interrogated, and $\tau$ is the averaging time. The advantage of optical clocks results from the optical clock frequency that can be up to five orders of magnitude higher than that of a microwave standard. There are two different types of optical atomic clocks. One uses radio-frequency traps to store ionized particles for the required long interrogation times, the other one employs optical lattices tuned to the so-called ``magic wavelength'' to confine neutral atoms. In order not to perturb the ionic clock transition, only a single ion or a few ions can be interrogated in the field-free region of the trap even though attempts are made to enlarge the number of ions in multiple ion trap clocks \cite{her12b}. The single ion trap clocks have therefore a higher instability according to Equation \ref{instability} as compared to neutral atom clocks where $N$ can easily approach or exceed $10^4$. On the other hand, particular optical transitions in ions like Al$^+$ \cite{cho10} are extremely insensitive to external perturbations e.g. the electromagnetic radiation of the ambient temperature field \cite{saf11}. It is therefore presently not clear which type of clocks will eventually be the best one for a re-definition of the second. Consequently, PTB investigates both approaches operating single ion Yb$^+$ and Al$^+$ (Section \ref{Single ion clocks}) and neutral atom Sr lattice clocks (Section \ref{Neutral atom clocks}). 
In Section \ref{Challenges} the most challenging problems of these state-of-the art clocks are discussed as well as the additional tasks undertaken to support the path towards a future re-definition of the second (Section \ref{Conclusions}).  

\section{Single ion clocks}
\label{Single ion clocks}
\begin{figure}[htbp]
\centering\includegraphics[width=7cm]{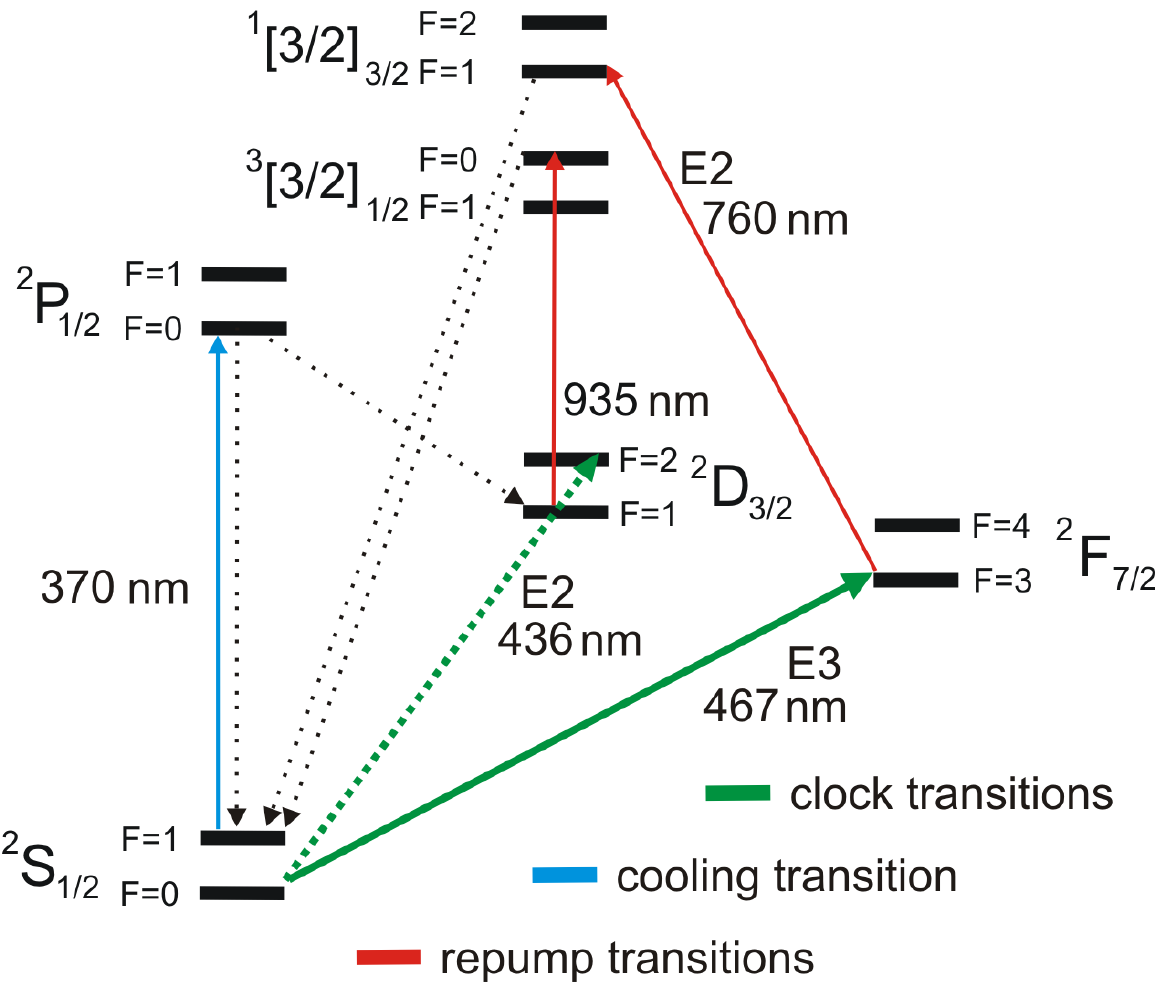}
\caption{\label{Yb-TermScheme} Energy level diagram of the relevant transitions for Yb$^+$ clocks}
\end{figure}

Single ion clocks employ an oscillating radio frequency field to trap the ion in a field-free center between the electrodes. Different electrode configurations are used in optical frequency standards like the Paul trap arrangement \cite{pau89}, the end cap trap \cite{sch93}, or the linear trap \cite{rai92a}. The first two arrangements lead to a point-like field-free region for only a single ion whereas the linear trap has a weaker confinement on a line between the linear electrodes that permits the storage of a chain of multiple ions. To use a transition with a small natural line width often a quadrupole transition 
(e.g. $^2$S$_{1/2}$ - $^2$D$_{5/2}$) is employed like in the Ca$^+$ \cite{mat08a,chw09,hua12}, Sr$^+$ \cite{mar04,mad12}, Hg$^+$ \cite{ros08}, or Yb$^+$ \cite{god14,hun14} clocks  (Fig. \ref{Yb-TermScheme}). All these transitions have a natural lifetime that supports a line width on the order of a Hertz.

\paragraph{Yb$^+$ Single ion clock}
Besides the $^2$S$_{1/2}$ - $^2$D$_{3/2}$ quadrupole transition in Yb$^+$, the $^2$S$_{1/2}$ - $^2$F$_{7/2}$ octupole transition with an extremely high natural line quality factor (Q $\approx 10^{23}$) has been used as a reference in an optical frequency standard at the National Physical Laboratory (UK) \cite{god14} and at PTB \cite{hun14}. Both, the quadrupole and octupole clock transitions and the transitions used for cooling and manipulation in Yb$^+$ are easily accessible by diode lasers. For the quadrupole transition, a fractional uncertainty of $1.1 \times 10^{-16}$ has recently been estimated, dominated by the contribution of the frequency shift due to thermal radiation at room temperature \cite{tam14}. 

Since the octupole transition is very weak, high irradiance of the probe laser is required to interrogate this clock transition therefore leading to a considerable ac Stark shift. Recently, a method has been devised \cite{yud10} and   
employed \cite{hun12} to cancel the light shift by a suitable interrogation sequence. The light shift is compensated 
by a detuning of the clock laser during the interaction pulses of the interrogation. The detuning is controlled by interleaved interrogations with single-pulse Rabi excitations. An additional $\pi$ pulse removes the linear dependence between the uncompensated light shift and the frequency of the central resonance feature leaving only a third-order dependence. With this method the light shift can be suppressed to the low $10^{-18}$ regime.
The presently largest contribution to the systematic uncertainty is the Stark shift caused by the ambient thermal radiation, which is in turn limited by the 50\% relative uncertainty in the static differential polarizability. A more precise measurement of this property using near-infrared laser radiation and a detailed study of the thermal environment in combination with the suppressed light shift of the probe laser promises a total uncertainty in the low $10^{-18}$ range. In addition to this low uncertainty, the octupole transition is known for its particularly high sensitivity to temporal variation of the fine structure constant $\alpha$. Repeated measurements of both clock transition frequencies versus caesium fountain clocks with improved accuracy over the last 7 years have recently confirmed the previous limit on $d\alpha/dt$ and provide the most stringent limit on a temporal variation of the proton-to-electron mass ratio \cite{god14,hun14}. With this exceptionally small uncertainty the Yb$^+$ octupole clock is one of the optical clocks with the smallest estimated uncertainties. As one of the next steps the frequencies of two independent Yb$^+$ clocks of PTB will be compared.


\paragraph{$^{27}$Al$^+$ quantum logic clock}
The optical $^{27}$Al$^+$ clock makes use of a transition ($\lambda = 267$~nm) between the $^1$S$_0$ and $^3$P$_0$ states with vanishing angular momenta $J=0$. Such transitions have no quadrupole shift and only the small nuclear magnetic moments contribute to the linear and quadratic Zeeman shift. The $^{27}$Al$^+$ clock transition can be reached with conventional technology but the transitions usually used for laser cooling and state detection are in the far ultraviolet. Wineland's group at NIST has devised a scheme where an auxiliary so-called {\it logic ion} (e.g. $^9$Be$^+$) provides sympathetic laser cooling, state initialization, and detection for a simultaneously trapped $^{27}$Al$^+$ clock ion. The NIST group has operated two such clocks with estimated fractional uncertainties in the $10^{-18}$ range and their frequencies agreed within the combined uncertainties \cite{cho10}. A number of institutes has started to set up their own optical $^{27}$Al$^+$ clock, among them PTB. In PTB's set-up  Ca$^+$ is used as the logic ion since the transitions can be addressed with diode lasers and cooling is sufficient \cite{wue12}. PTB's $^{27}$Al$^+$ clock will be used as a transportable clock for comparisons of remote optical clocks and for novel applications e.g. together with fiber links in relativistic geodesy experiments. 

\section{Neutral atom clocks}
\label{Neutral atom clocks}
Neutral atomic clocks have become competitive to the best ion clocks again only after the invention of Katori \cite{kat03} that the huge light shift of the atomic levels used for the clock transition can be made equal in a magic wavelength optical dipole trap. The basics are described elsewhere in this Special Issue and need not be considered here in more detail.        

\begin{figure}[htbp]
\centering\includegraphics[width=7cm]{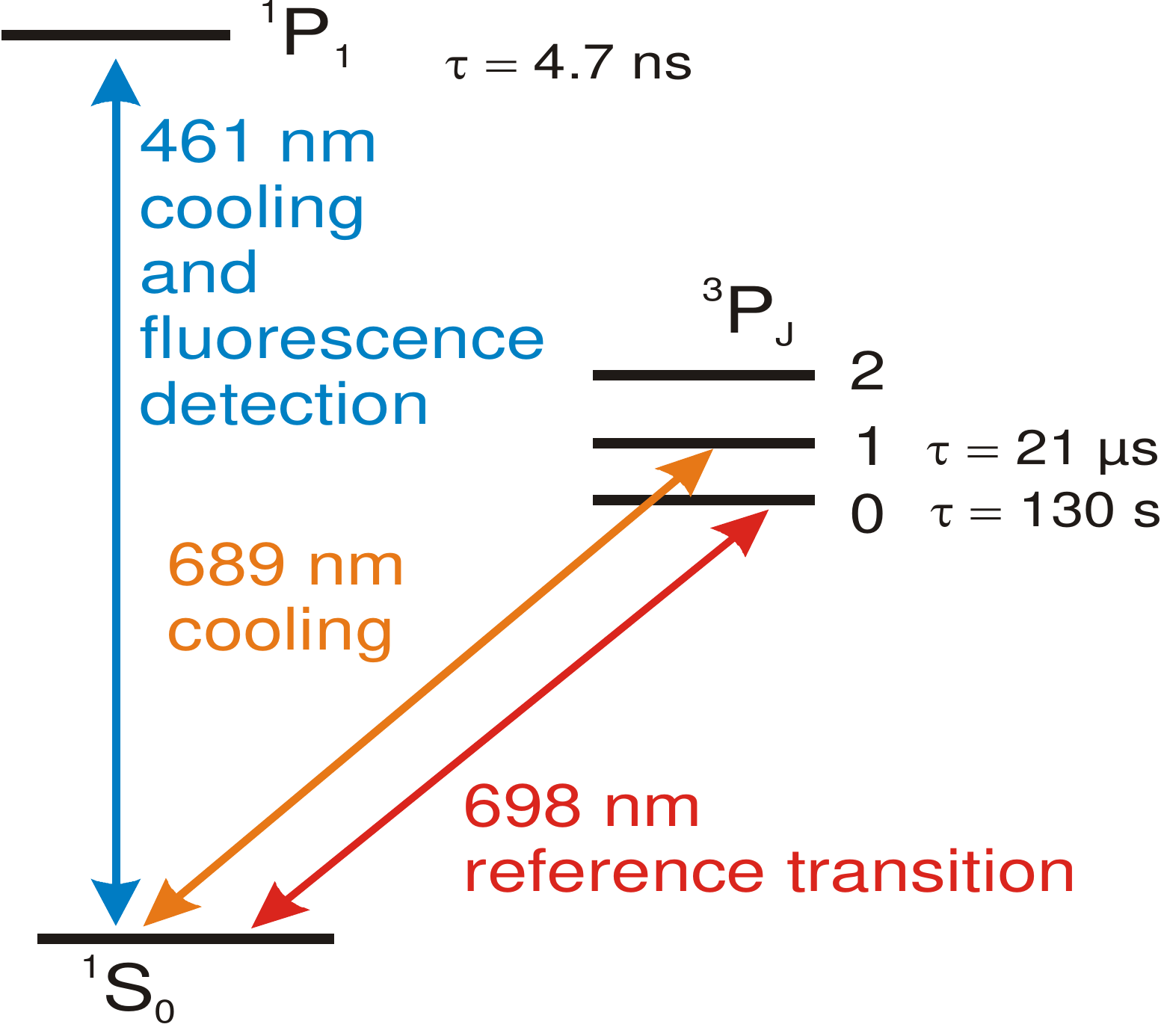}
\caption{\label{Fig: SrTermScheme} Energy level diagram of the relevant optical transitions in the Sr optical atomic lattice clock}
\end{figure}

\paragraph{PTB's Sr lattice clock}
\label{SrLatticeClock}%
From all optical atomic lattice clocks employing e.g. Sr, Yb, or Hg, the Sr lattice clock is currently the one that has found most widespread use either already in operation (see e.g. \cite{ush14,blo14,let13,fal14,aka14,yam11}) or under investigation. The clock transition $\lambda$ = 698~nm between the ground state $^1S_0$ and the $^3P_0$ state (see Fig. \ref{Fig: SrTermScheme}) is weakly allowed in the $^{87}$Sr isotope with nuclear spin $F=9/2$ with an excited state life time of about 130~s due to the hyperfine interaction. 

The Sr atoms are first cooled on the $^1$S$_0 - ^1$P$_1$ transition with $\lambda$ = 461~nm and in a second stage on the 
$^1$S$_0 - ^3$P$_1$ transition before being trapped in an optical lattice operated at the magic wavelength of $\lambda$ = 813.4~nm. PTB has built and operates a Sr lattice clock \cite{fal11,mid12a,fal14} that has a similar physics package as the clocks in other institutes. The fractional instability of PTB's Sr stationary lattice clock is currently $ \sigma (1\;{\rm s}) = 2.6 \times 10^{-16}$ and falls off like $\tau^{-1/2}$ into the upper $10^{-18}$ range. The fractional uncertainty of this clock was estimated to be $3 \times 10^{-17}$ \cite{fal14}. Even lower estimated standard uncertainties at the $10^{-18}$ level have been derived for a clock based on Sr atoms kept in a room temperature environment \cite{blo14} or in a cryogenic housing \cite{ush14}.


\begin{figure}[htbp]
\centering\includegraphics[width=10cm]{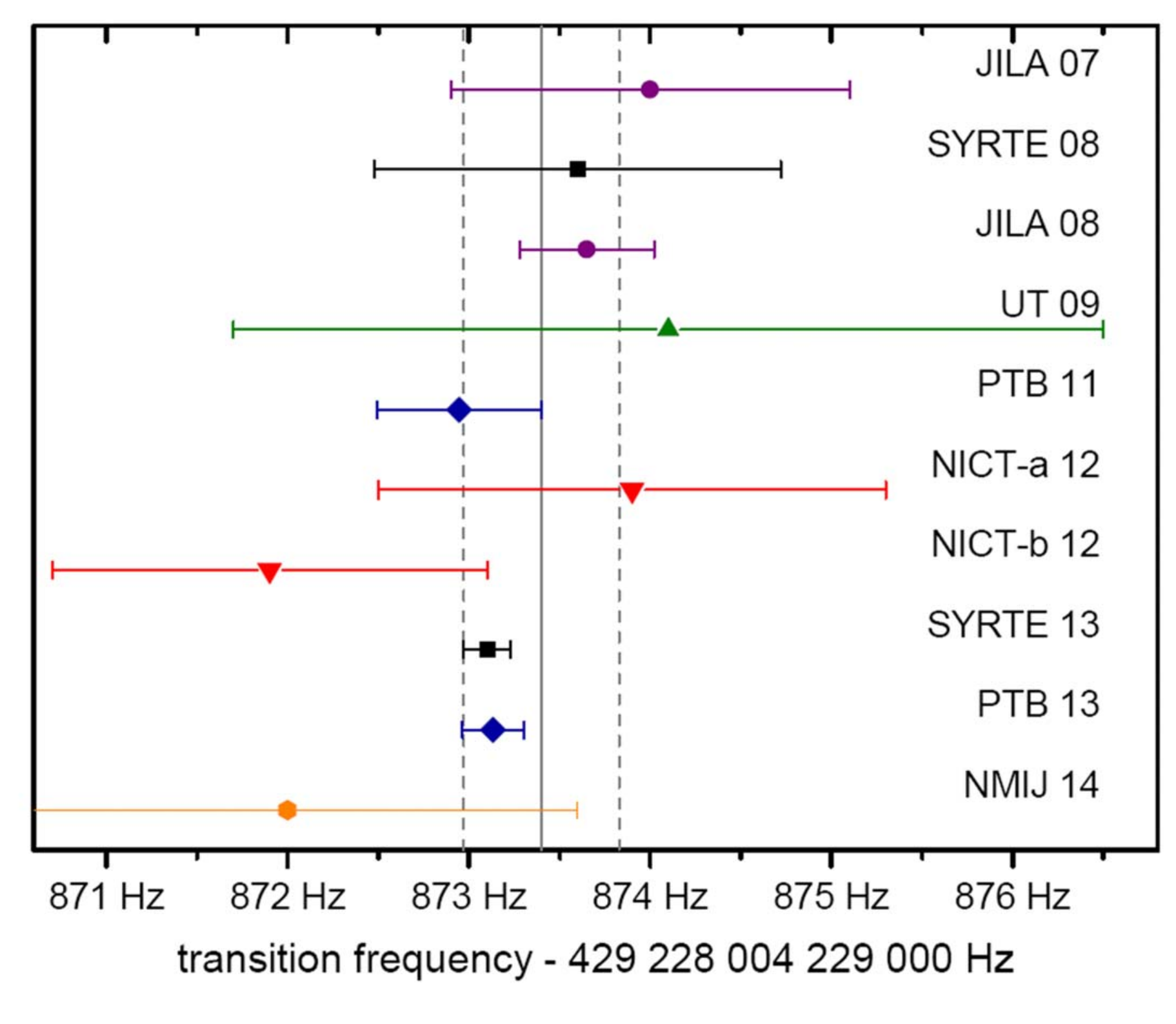}
\caption{\label{Fig:frequencies} Frequency measurements wrt primary caesium atomic clocks. JILA07: \cite{boy07}; SYRTE08: \cite{bai08}, JILA08: \cite{cam08}, UT09: \cite{hon09}, PTB11: \cite{fal11}, 
NICT-a 12: \cite{yam12}, NICT-b 12: \cite{mat12}, SYRTE 13: \cite{let13}, PTB13: \cite{fal14}, NMIJ14: \cite{aka14}. The full vertical line indicates the value for use as a secondary representation of the second with the estimated uncertainty (dashed lines) recommended by the CIPM \cite{cip12}.} 
\end{figure}

A considerable number of frequency measurements of the $^{87}$Sr optical lattice clocks with respect to primary caesium clocks has been performed over the past years either by local comparisons in the same institute or by linking the optical standards to remote clocks via fibers or satellites (Fig. \ref{Fig:frequencies}). The upper six measurements displayed in Fig. \ref{Fig:frequencies} have been used to recommend the frequency of this transition as $\nu_{87Sr} = 429~228~004~229~873.4$~Hz with a relative standard uncertainty of $1 \times 10^{-15}$ estimated for the secondary representation of the second. The four most recent measurements fall into the estimated uncertainty band but seem to tend towards a lower frequency. From the total of the measurements displayed here the CIPM might recommend at its next meeting a slightly lower value and also slightly reduce the estimated uncertainty of this recommendation since the larger number of independent measurements with reduced uncertainties gives a higher confidence to the estimated value.

\paragraph{Transportable Sr lattice clock}

\begin{figure}[htbp]
\centering\includegraphics[width=13cm]{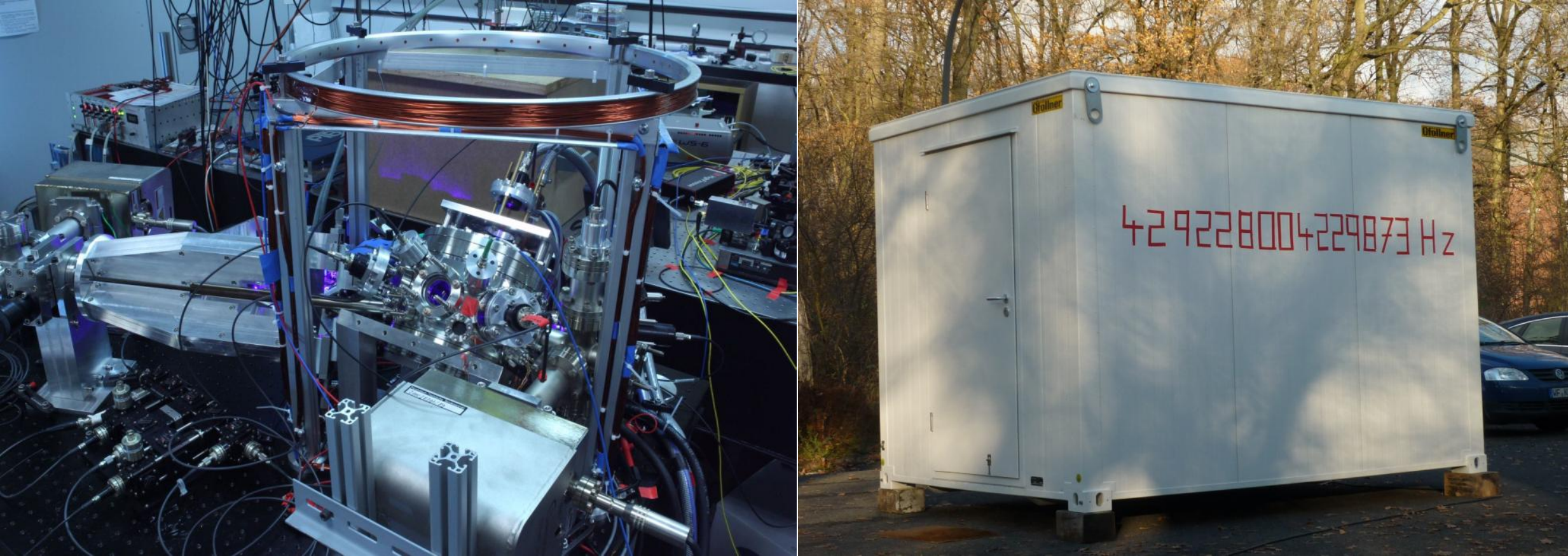}
\caption{\label{Fig:TransportableLatticeClock} Physics package of PTB's transportable lattice clock (left) and container for the transportation of the clock (right).} 
\end{figure}

Today, the estimated optical frequencies of clocks like the ones presented in sections \ref{Single ion clocks} and \ref{Neutral atom clocks} can only be compared with such low uncertainties via optical fibers \cite{pre12} or free-space links. Therefore attempts are made in several projects or institutes to set up transportable optical clocks that are capable to be used for clock comparisons with appropriate low uncertainties. A multinational consortium sponsored by the European Union Seventh Framework Program \cite{pol14} is developing a transportable standard based on a Sr lattice clock which later also could be used for space applications \cite{bon15}. PTB has set up a transportable Sr lattice clock (Fig. \ref{Fig:TransportableLatticeClock}) to be used for comparisons of remote optical clocks and for novel applications e.g. in relativistic geodesy. In contrast to the two other approaches this transportable clock is not aiming towards ultimate compactness but will be operated in a transportable container with stable environmental conditions at any place. The transportable Sr lattice clock of PTB is in principle not very different from that of the stationary clocks with a few specific properties \cite{vog15}. The diode laser systems have been designed for stability and compactness and are similar to the ones described in \cite{bon15}. The transportable standard uses a Zeeman slower with permanent magnets for decelerating the Sr atoms towards the magneto-optic trap in order to reduce the electric power of the system. Since one of the  largest contributions to the uncertainty of the Sr lattice clocks results from the black body shift 
care was taken to achieve a homogeneous temperature environment of the vacuum chamber that can be correctly measured. A low uncertainty is achievable since the sensitivity coefficient is very well established \cite{mid12a}.

First measurements with the transportable clock show a fractional instability of $1 \times 10^{-16}$ at 500~s. Its frequency agrees with that one of the stationary standards within the current estimated uncertainty of $10^{-15}$.  From the evaluation of the uncertainty budget very similar to the one presented in \cite{fal14} a final fractional uncertainty in the low $10^{-17}$ region is expected. After the final evaluation the transportable clock will be used for comparisons of remote optical clocks and for novel applications e.g. together with fiber links in relativistic geodesy experiments.

\section{Challenges to be met on the way to a re-definition of the second}
\label{Challenges} 
With the current status of the optical clocks (Fig. \ref{OptvsMicro}) as outlined above the question is often asked {\it when and how a re-definition of the second might take place}. P. Gill \cite{gil11} has addressed this question a few years ago and it is timely to revisit his arguments and complement them. In trying to answer this question it is helpful to ask two other questions: {\it what are the requirements to be met for a re-definition?} and {\it who will benefit from a re-definition of the second?} In the next paragraphs I first shall concentrate on the technical challenges to be met as prerequisites for answering the first question.

\paragraph{Ultra-stable lasers for interrogation of the clock transition}
\label{StableLasers}
The achievable quality of an optical clock depends crucially on the stability of the laser used to interrogate the clock transition. The investigation of all the systematic effects limiting the achievable uncertainty in the $10^{-18}$ regime and below can be achieved only in a reasonable time if the instability of the interrogating laser is smaller than the instability allowed for by interrogating the clock transition. 

\begin{figure}[htbp]
\centering\includegraphics[width=10cm]{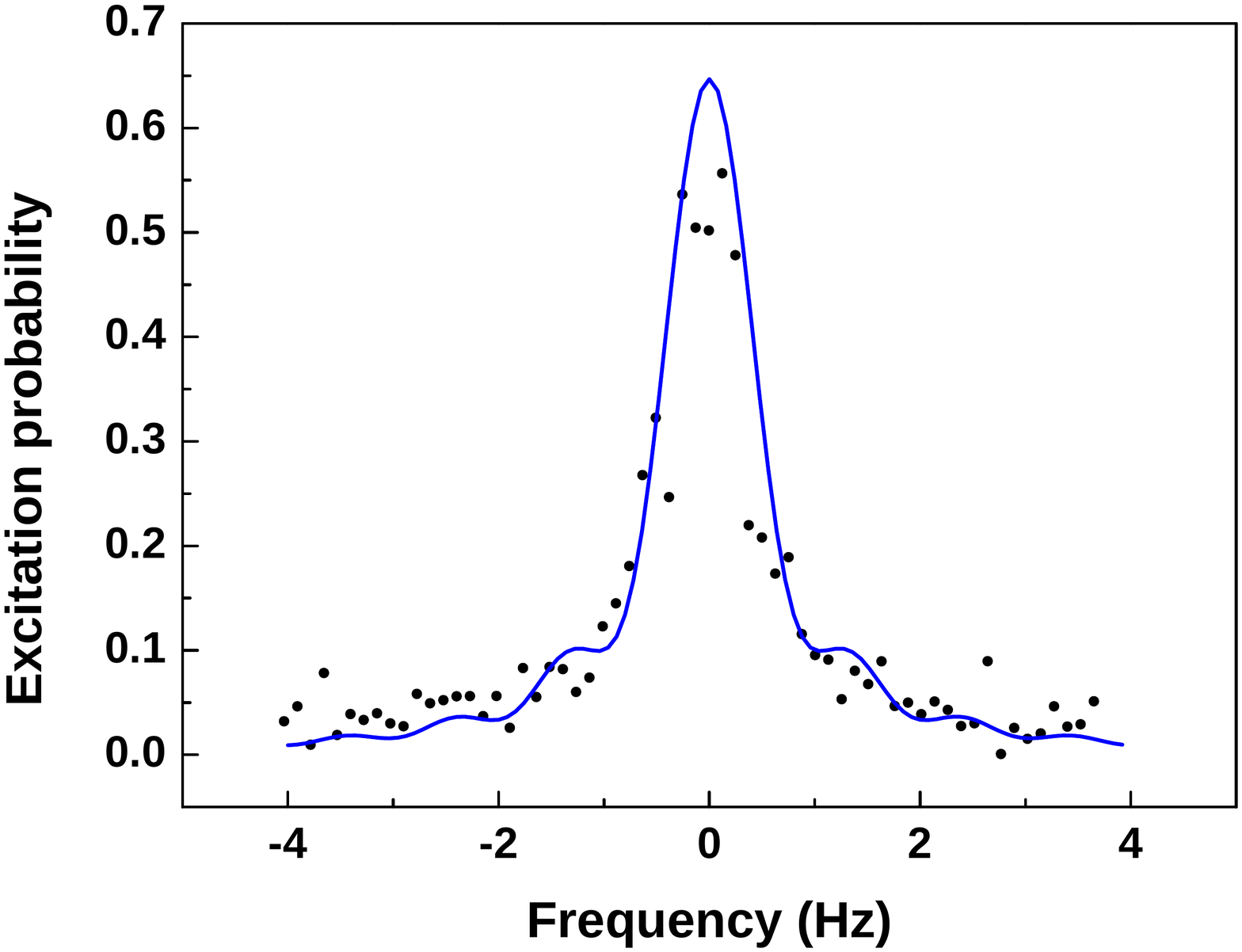}
\caption{\label{SrLine} Measured 0.9~Hz line width of the Sr clock transition using a 50 cm long ULE resonator stabilized laser \cite{hae15}.}
\end{figure}

Currently three different techniques are used or are under investigation to obtain super-stable lasers. The most widespread scheme relies on the stabilization of the laser frequency to a high-finesse and super-stable optical Fabry-P\'erot resonator by means of the Pound-Drever-Hall frequency stabilization scheme \cite{dre83}. 
An interesting alternative to this approach has recently been investigated where the laser is stabilized to spectral holes burnt into an ensemble of Eu dopant ions in a solid at cryogenic temperatures \cite{che11,lei13,tho13}. Another method with high potential has been proposed that uses an active light source based on having alkaline-earth atoms in an optical lattice. These atoms can collectively emit photons on an ultra-narrow clock transition into the mode of a high Q resonator \cite{mei09}.
PTB's clock lasers use a pre-stabilization to a Fabry-P\'erot interferometer. High finesse mirrors optically contacted to a spacer made from a material with low thermal expansion coefficient at the operating temperature like Ultra Low Expansion glass (ULE) or silicon \cite{kes12} are employed. To reduce the influence of ambient vibrations on the optical path length between the mirrors and, hence, on the eigen frequencies of the resonator, a number of provisions has to be made. Different ways to achieve a vibration insensitive mounting of the resonator have been realized (see e.g. \cite{mil09} and references therein). Since PTB operates different optical clocks with different optical frequencies several ultra stable optical resonators are in use that can be compared and together with femtosecond optical frequency combs used for connecting the different clocks \cite{hag13}. With the best resonators line widths around one Hertz (Fig. \ref{SrLine}) are readily obtained when narrow transitions like in the Sr lattice clock (Section \ref{SrLatticeClock}) are interrogated. This line width is Fourier limited due to the short interaction time of about 1~s and it has been shown that a special laser could support a line width of 40~mHz (at 1.5~$\mu$m) \cite{kes12}. However, to resolve the natural line width of about 8~mHz in the Sr clock challenging improvements of the pre-stabilized lasers are needed to reduce their phase fluctuations necessary for the required long interrogation times. 

\paragraph{Comparisons of optical clocks remote from each other}
\label{Methods to compare distant optical clocks}
The frequent comparisons of clocks that are necessary to maintain the International Atomic Time Scale TAI and the International Coordinated Time Scale UTC rely on well established technologies using radio frequency signals from and via satellites (see e.g. \cite{bau15} in this Special Issue). The frequencies of two Sr lattice clocks between Japan and Germany were compared via two-way carrier phase time and frequency transfer using a geostationary satellite  in a first demonstration \cite{hac14}. The fractional uncertainty of $1.6 \times 10^{-15}$ for a measurement time of one day shows that the currently available standard techniques are not capable to support the accuracy and instability of optical clocks in the $10^{-18}$ range. With a larger transponder band width this situation could be improved. The {\it Atomic Clock Ensemble in Space} (ACES) project \cite{cac11} planned for the International Space Station after 2016 will make use of a microwave link that should lead to a fractional instability and inaccuracy of $3 \times 10^{-16}$ for a one day measuring time. Optical links, either free space \cite{sam14} or via optical telecommunication fibers \cite{pre12} are much more promising. On a fiber link between PTB and the Max-Planck Institute of Quantum Optics at Garching near Munich it could be shown that over 920~km a fractional frequency instability (modified Allan
deviation) of $5 \times 10^{-15}$ in a one second integration time could be achieved reaching $10^{-18}$ in less than one hundred seconds. At the time of this writing, PTB and SYRTE are setting up a continuous fiber link between Braunschweig and Paris to allow for frequency comparisons of their optical clocks during the next years. Similar links e.g. between NPL and SYRTE, in Italy and at several other places are currently also implemented. This topic is addressed elsewhere in this Special Issue. To conclude this paragraph, dedicated fiber links are capable and used already to compare optical clocks on a continental scale without compromising the performance of the best optical clocks. However, to intercompare the frequencies of the optical clocks on the $10^{-18}$ level a very accurate knowledge of the difference in the gravitational potential at the locations of the clocks corresponding to a cm hight level would be necessary that is currently not yet available. This requires challenging improvements in the geodetic modeling. For the next years therefore transportable standards are necessary to achieve this goal and seem furthermore to be the only choice to compare the best optical clocks on an intercontinental scale.

\paragraph{Frequency ratios of optical clocks}
To evaluate and validate the different optical clocks with the highest accuracy it will be necessary to compare these clocks next to each other at the same gravitational potential and without making a detour via the microwave clocks. The measurement of direct optical frequency ratio between different optical standards in the same laboratory will allow one to overcome the aforementioned limitations since such a ratio is independent from the frequency of the caesium clock. The use of such ratios promises a means to evaluate the combined uncertainty of the optical standards with an accuracy limited by the standards themselves. In the simplest case two independent clocks of the same type are operated in the same laboratory in the $10^{-18}$ regime like in the case of the $^{27}$Al$^+$ clock \cite{cho10} or the Sr lattice clock \cite{ush14}. 
The numerical value of frequency ratios measured between optical clocks whose frequencies are very different can easily be transported to another laboratory which uses the same clock combination. As an example consider the ratio of the frequencies of the Yb$^+$ octupole and quadrupole standards that are investigated both at PTB \cite{hun14} and NPL \cite{god14}. The comparison of the frequency ratios measured in both institutes with high accuracy will give a means to evaluate and validate the estimated uncertainties. Another example would be the ratio between the frequencies of the Yb and Sr neutral lattice standards operated at NIST \cite{hin13} and JILA \cite{blo14} in Boulder, USA, or in the Tokyo metropolitan area. Furthermore, if enough such frequency ratios between clocks with optical frequencies $\nu_A$, $\nu_B$, and $\nu_C$ are available they could be combined to make a closure like e.g.
\begin{equation}
\frac{\nu_A}{\nu_B} \times \frac{\nu_B}{\nu_C} \times \frac{\nu_C}{\nu_A} = 1 ?
\label{FrequencyRatios}
\end{equation}
The joint working group WGFS of the CCL and the CCL of the CIPM are currently developing methods and procedures to collect and make use of such data to boost the confidence into the particular measurements from the consistency of the ensemble. 

\paragraph{Supporting developments for utilization and novel applications of optical clocks}
The prospects and progress of optical atomic clocks has attracted considerable attention to achieve higher accuracy in  fields where microwave atomic clocks are used but also for the development of novel research and application fields.
The current drivers for better optical atomic clocks are mainly basic research, time and frequency metrology and novel applications. The search for better limits for the constancy of the constants \cite{god14, hun14}, secondary representations of the second, and a future {\it Relativistic Geodesy} represent one of a few examples from each of these applications. As it is well known according to General Relativity, two optical clocks in a different gravitational potential show different frequencies \cite{cho10a} when compared and this effect is taken into account for a long time for the comparison of atomic clocks in the international time scales TAI and UTC. With a difference of the fractional frequency $\delta \nu / \nu \approx 10^{-16}$ per meter height difference near the surface of Earth, the expected $10^{-18}$ fractional uncertainty of the best optical clocks requires either the determination of the exact gravitational potential equivalent to 1~cm in height or allows one to determine the difference in the gravitational potential at the locations of the two clocks. The potential of such a Relativistic Geodesy has been recognized a long time ago \cite{bje85} but it now can be realized provided that suitable transportable clocks and fiber links are available. 

A wider use of optical clocks in this and other fields will depend heavily on a higher degree of compactness, reliability, and increased operation time and commercial availability. Even though the basic building blocks for such clocks are currently developed, e.g. for the transportable optical clocks or for space applications, there are further engineering steps required. Optical clocks for special applications like Relativistic Geodesy and other ground and space applications will then immediately find the entrance into the well established fields where atomic clocks are used today, e.g. for time keeping and all other fields of frequency metrology. 

\section{Conclusions and possible roadmap} 
\label{Conclusions}
The status of the optical atomic clocks reported here at the example of the work at PTB and others shows that the increase in accuracy and stability continues with unbroken pace thereby leaving behind more and more the best realization of the definition of the second in the SI - the caesium atomic clock. In preparation for a future re-definition of the second in terms of an optical atomic clock several issues have been identified that are necessary or desirable prerequisites. Let us now come back to the question {\it who will benefit from a re-definition of the second?}. Due to the already very low uncertainty of the primary caesium clocks approaching $10^{-16}$ it seems safe to conclude that the present caesium definition will serve for some time most of industry's needs and the secondary representations will serve most of the necessities in science. In the planned revision of the SI in 2018 \cite{NewSI} ``four of the SI base units -- namely the kilogram, the ampere, the kelvin and the mole -- will be redefined in terms of constants; the new definitions will be based on fixed numerical values of the Planck constant ($h$), the elementary charge ($e$), the Boltzmann constant ($k_\mathrm{B}$), and the Avogadro constant ($N_\mathrm{A}$), respectively''. After this revision, all base units, except for the mole, will essentially be derived from the value of the second in the SI. Since this realization at present is several orders of magnitude more accurate as needed for the other SI base units there is no direct need to re-define the second under this aspect. On the other hand the new SI will not be affected in any negative way by a soon re-definition of the second. The cost associated with a re-definition would be very low since the caesium technology would be still available and used in all applications where the current accuracy is sufficient.

After a re-definition the generation of the time scales would benefit directly by the available more stable and more accurate clocks that could be included into the generation of TAI and UTC and replace the caesium clocks gradually. The first steps along this line have been taken already \cite{let13,wol06}. A re-definition in conjunction with optical telecommunication fiber networks allows the much higher accuracy and stability to be delivered to virtually all places where phase-coherent fiber links are available or can be installed. The short cycle times for exchange in the fiber networks for improved efficiency will allow such an installation of phase-coherent components step by step on a broad scale. Such an opportunity will boost novel applications and services from the very beginning. 

From these points of view an early re-definition is possible and very desirable. There are, however, considerations that ask for a little more patience. There are currently a number of different optical clocks with the prospects of further improvement. Currently there is no definite argument for a decision between atomic and ionic species and in both classes there are several promising candidates for a re-definition of the second. It can be expected that the stunning pace in the development of optical atomic clocks seen in Fig. \ref{OptvsMicro} will at some time slow down and could reduce the large number of the present candidates. On the other hand there might be novel concepts with prospects to achieve e.g. a reduced interaction of the clock transition with all sorts of environmental perturbations. A transition between two isomeric states in $^{229}$Th has attracted much consideration (see e.g. \cite{pei15} in this Special Issue) for such a possibly better isolated clock transition. PTB has a project \cite{pei03} to find this clock transition via bridge processes \cite{por10b} in ion clouds and in crystals doped with $^{229}$Th. One should also keep in mind that many of the groups contributing to the rapid progress in optical clocks are motivated by questions from basic science, like performing improved tests  of fundamental principles in quantum physics and relativity. Many of these studies have benefited from the available variety of different atomic clocks, with different sensitivities to specific interactions. It therefore may be detrimental to scientific progress to narrow the choice of investigated systems prematurely as long as the current pace goes on with unbroken speed (Fig. \ref{OptvsMicro}).  

Eventually a consensus has to be found for the most suitable candidate. The achievable uncertainty is one but not the only one criterion. There are also other arguments for particular candidates as the prospects of the future advancement or the ease of operation and use. With all these issues in mind we are in the position to develop an updated roadmap (similar to the one published by Euramet \cite{roa12}) that could lead to a re-definition of the second by an optical atomic clock within the next two decades.  
       
\vspace{1cm}
{\bf Acknowledgment:}
The author acknowledges very valuable discussions with E. Peik, N. Huntemann, H. Schnatz, Ch. Lisdat, U. Sterr, S. Vogt, S. H\"afner, P. O. Schmidt who contributed munificently their results. I also acknowledge support by the DFG RTG 1729 ``Fundamentals and applications of ultra-cold matter''. Very helpful discussions with the members of the CCL-CCTF Frequency Standards Working Group are gratefully acknowledged. I am indebted to J. Frieling for his careful reading leading to this new version. 






\end{document}